\documentclass[12pt,twoside,authoryear]{article}

\usepackage[hidelinks]{hyperref}
\usepackage{authblk}
\usepackage{verbatim}
\usepackage{siunitx}
\usepackage{graphicx}
\usepackage[natbib,authordate]{biblatex-chicago}
\usepackage{tikz}
\usepackage{setspace}
\usepackage[font={small},labelfont=bf]{caption}

\usepackage[margin=1in,includefoot,footskip=30pt]{geometry}

\title{Gerrymandering and Compactness: \\ Implementation Flexibility and Abuse}

\author[a,b,c]{Richard Barnes}
\author[c]{Justin Solomon}

\affil[a]{Energy \& Resources Group, UC Berkeley, Berkeley, CA, 94720. ORCID: 0000-0002-0204-6040}
\affil[b]{Berkeley Institute for Data Science, UC Berkeley, Berkeley, CA, 94720.}
\affil[c]{Computer Science and Artificial Intelligence Laboratory (CSAIL), MIT, Cambridge, USA, 02139}

\date{}

\begin{document}

{\noindent\footnotesize\raggedright Please cite: Barnes and Solomon (2020). Gerrymandering and Compactness: Implementation Flexibility and Abuse. Political Analysis. doi: 10.1017/pan.2020.36}

{\let\newpage\relax\maketitle}

\begin{abstract}
\noindent Political districts may be drawn to favor one group or political party over another, or \textit{gerrymandered}. A number of measurements have been suggested as ways to detect and prevent such behavior. These measures give concrete axes along which districts and districting plans can be compared. However, measurement values are affected by both noise and the compounding effects of seemingly innocuous implementation decisions. Such issues will arise for any measure. As a case study demonstrating the effect, we show that commonly-used measures of geometric compactness for district boundaries are affected by several factors irrelevant to fairness or compliance with civil rights law. We further show that an adversary could manipulate measurements to affect the assessment of a given plan. This instability complicates using these measurements as legislative or judicial standards to counteract unfair redistricting practices. This paper accompanies the release of packages in C++, Python, and R that correctly, efficiently, and reproducibly calculate a variety of compactness scores.
\end{abstract}

Gerrymandering is the practice of designing political districts whose shapes serve some agenda, often the consolidation of power by a political party or the disenfranchisement of a group such as a minority population. In 2018, litigation relating to gerrymandering was underway in at least twelve U.S.\ states, with several cases reaching the U.S.\ Supreme Court. In the same year, Colorado, Michigan, Missouri, Ohio, and Utah approved referendums intended to limit gerrymandering through the use of independent commissions. In 2019, cases from Maryland and North Carolina reached the Supreme Court, which ultimately rejected the federal judiciary's role in districting. With this decision both major political parties have begun to focus on state-level legislation and legal proceedings. The high-profile nature of these cases and citizens' demands for solutions has led to interest in developing ways to measure district fairness.

Such measures give concrete axes along which districts and districting plans can be compared. However, the data used to measure a district may have noise or errors. The choices surrounding how a measurement is made also interact with each other to form a ``garden of forking paths"~\citep{Gelman2013} in which each choice affects the outcome of the others. This compounding can have a significant effect on certain scores that appear mathematically reasonable. We demonstrate this issue in a case study by showing that common ways of measuring the shape, or \textit{compactness}, of districts are affected by several factors irrelevant to fairness or compliance with civil rights law. We further show that an adversary could actively manipulate these scores to affect the assessment of a given plan.

The U.S.\ Supreme Court has considered the shape of electoral districts in a number of cases including
Reynolds v.\ Sims (1964),
Gaffney v.\ Cummings (1973),
Thornburg v.\ Gingles (1986),
Shaw v.\ Reno (1993),
Bush v.\ Vera (1996),
Karcher v.\ Daggett (1983), and
Cooper v.\ Harris (2017). Aside from legal precedents, 37 states require that their state legislative districts be compact and 18 explicitly require compactness of their congressional districts.

Mathematically, the compactness of a district is a geometric quantity intended to capture how ``contorted" or ``oddly shaped" a district is. Although compact districts can also be gerrymandered and contorted shapes can arise from geographic or legal necessity, such as rivers or municipal boundaries, poor geometry is often understood as a signal of gerrymandering. For instance, in Bush v.\ Vera (1996), the Supreme Court condemned districts that were ``bizarrely shaped and far from compact." For these reasons, compactness is quantified during redistricting, though Thornburg v.\ Gingles (1986) demonstrates that many other considerations must also be made.

Many measures of compactness exist~\citep{altman1998,Niemi1990,chambers2010}, and mathematicians and legislators continue to debate their relative merits in promoting desirable district shapes. There has been less discussion, however, about how compactness scores should be implemented in practice.

Here, we use the US Census Bureau's 2015 Cartographic Boundary and TIGER/Line data~\citep{shapefiles} to show how the variables used to calculate compactness are complicated by reality and how, even once quantitative scores are defined, confounding factors including geography, topography, cartographic projections, and resolution complicate implementation. Together, the ambiguities we expose provide a high degree of \emph{flexibility}. We show that this flexibility can be exploited to engineer compactness scores that allow convoluted and gerrymandered districts to meet quantitative standards designed to prevent such abuse.

If policymakers are unaware that quantitative measures of electoral district fairness may be both intentionally and unintentionally manipulated to give a variety of outcomes, they may push to enact standards that are either insufficient or that can be gamed. This problem arose as litigants sought to use the efficiency gap~\citep{Stephanopoulos2014} to detect gerrymandering even as it was shown that the measure was problematic~\citep{Alexeev2017,Bernstein2017,Chambers2017,Veomett2018}. Here, we show that similar problems exist for compactness. We also suggest that implementation flexibility and the accompanying potential for abuse is a general property of trying to quantify electoral fairness.

Section \ref{sec:considerations} is technical and exposes the full complexity and consequences of the many considerations that must go into calculating aspects of a compactness measurement. Section \ref{sec:results} provides a non-technical summary of the results and shows how the methods we discuss here can be abused. Section \ref{sec:discussion} concludes with recommendations for the development and fair characterization of compactness scores. We additionally provide a model software implementation intended to avoid the pitfalls we highlight. All of the examples presented and some of the terminology used stem from United States geopolitics, but our ideas are applicable to districts in any context. Although we focus on compactness, our work provides a cautionary case study revealing challenges in quantifying any measure of gerrymandering.

\section{Technical Considerations}
\label{sec:considerations}

In this section we determine how compactness is affected by
(1)~the choice of mathematical definition,
(2)~contiguity,
(3)~topological holes,
(4)~the boundaries of political superunits,
(5)~map projection,
(6)~topography,
(7)~data resolution,
(8)~floating-point calculations,
and
(9)~whether alternative choices were possible in drawing a district's boundaries.
We determine how often issues arise and quantify the impact of each of these considerations on measures of compactness. In Section \ref{sec:results}, we will quantify the net impact when all of these considerations are combined, finding that each makes at least some contribution to affecting the quality of the measurements.

\subsection{Definitions of Compactness}

We identified over 24 different measures of compactness in the literature~\citep{altman1998,Niemi1990}.
Of these, we consider three of the most widely-used and their variants. These are illustrated in \autoref{fig:interps} and are as follows:
\begin{enumerate}
\item Polsby--Popper~\citep{polsby1991}: Given as $4\pi A/P^2$ where $A$ is the area of a district and $P$ its perimeter. This score is also known as the ``isoperimetric ratio"~\citep{deford2018total}.
\item Reock~\citep{reock1961}: the ratio of a district's area to the area of its minimum bounding circle. Finding this circle is non-trivial; an efficient algorithm and associated implementation is given by \citet{gartner1999}.
\item Convex Hull~\citep{Niemi1990}: the ratio of a district's area to the area of its convex hull, the minimum convex shape that completely contains the district.
\end{enumerate}
All of the above scores are in the range $[0,1]$ with higher values indicating greater compactness. Low values may indicate potential gerrymandering.

These scores are purely geometric. It may be that scores incorporating population densities or other demographic data provide a better means of measuring gerrymandering \citep{Niemi1990,Eig1981}, but they are outside the scope of our experiments. Regardless, all scores are subject to implementation flexibility of some sort. In fact, incorporating additional data might even \emph{exacerbate} the issues we discuss, since doing so would create additional opportunities for implementation flexibility.

\subsection{Data}

In our experiments, we draw geographic information from the US Census Bureau's 2015 Cartographic Boundary and TIGER/Line data~\citep{shapefiles} and use it to explore how implementation choices affect the measurement of the electoral districts of the 114th U.S.\ Congress. The Bureau's data comes in several different scales or resolutions: 1:500,000 (500k), 1:5,000,000 (5m), and 1:20,000,000 (20m). \autoref{fig:polysimp} depicts data at these different resolutions. High-resolution data (e.g., 500k) capture greater geographic detail at the expense of higher collection, storage, and computation costs whereas lower-resolution data (e.g., 20m) capture less geographic detail while reducing costs.

\subsection{Nomenclature}
\begin{figure}[h!]
\centering
\includegraphics[width=0.5\textwidth]{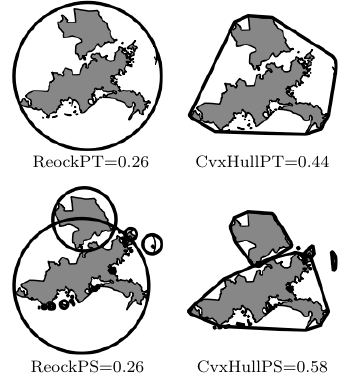}
\caption{Reock and Convex Hull scores for Louisiana 01 shown with both the \textit{PS} and \textit{PT} interpretation depicted. It is coincidental that ReockPT and ReockPS are the same here. Note that for both the ReockPT and CvxHullPT scores the hull polygons overlap; this overlap is potentially problematic since it could be considered double-counting. \label{fig:interps} }
\end{figure}

All of the measures we consider assume that an electoral district is described by a single planar polygon, without any holes. This assumption is problematic and leaves the measures under-specified. In reality, districts, such as those with islands (see \autoref{fig:interps}), are often comprised of many polygons. While holes in districts are rarer, they also can occur. We have to modify the scores so they can cope with reality, but there are many ways we can do this.

We will indicate whether or not contiguity is accounted for in a score by the suffixes \textit{PT} (polygons together) and \textit{PS} (polygons separate). Whether or not holes are accounted for will be indicated by the suffixes \textit{AH} (add holes) and \textit{SH} (subtract holes). If there is ambiguity regarding whether area, perimeter, or some other quantity is being treated in this way, then terms such as \textit{PTaSHp} (treat the area of the polygons together, subtract the perimeter of holes) may be used. The suffix \textit{B} indicates that a score accounts for constraints imposed by the boundaries of political superunits.

\subsection{Non-Contiguous Districts}
\begin{figure}[!h]
\centering
\includegraphics[width=0.3\columnwidth]{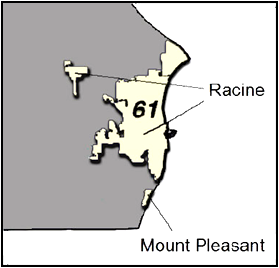}
\caption{Wisconsin's 61st Assembly District showing non-contiguous regions. See text for discussion. Figure drawn from~\citep{wisconsin}. \label{fig:wisconsin} }
\end{figure}

There is no federal requirement that districts must be contiguous and many states do not require it. Yet, most compactness measures assume contiguity. There are many ways of incorporating non-contiguity into compactness sores and each has a large effect. Avoiding the issue by requiring contiguity is likely impossible. Islands, such as Hawaii, make districts non-contiguous unless large bodies of water are included in the district, as discussed below. Disconnected districts may also arise in other ways. Civil rights considerations have given Louisiana 01, depicted in \autoref{fig:interps}, two large portions separated by Louisiana 02; Louisiana 02 was drawn as a majority-minority district following the passage of the Voting Rights Act of 1965. Wisconsin's 61st Assembly District (\autoref{fig:wisconsin}) exemplifies a different situation. The city of Racine, WI, became non-contiguous by annexing a nearby parcel, but both pieces of the city were included in the same district~\citep{altman2011}. For the 114th Congress 1:500,000 resolution data, 85 of 441 districts are not contiguous. Of the non-contiguous districts, the largest numbers of subdivisions were 580 (Alaska), 134 (Maine 02), 103 (Michigan 01), and 92 (Florida 26); the median was 5. Seventeen of the non-contiguous districts have portions that are separated by land; these include Kentucky 01 and Louisiana 01 (see \autoref{fig:interps}). 

The way we treat non-contiguous districts has significant effects on many proposed ways of measuring gerrymandering, including compactness scores. Treating the district as a single unit by, e.g., enclosing it in a single convex hull, will tend to result in lower compactness scores.
Treating the district as separate units and summing the areas of the units' enclosing hulls will result in higher compactness scores.

Mathematically speaking, although Polsby--Popper is usually calculated as $4\pi\frac{A}{P^2}$, there are several possibilities for extending this formula to non-contiguous districts, in particular $4\pi\sum_i^n \frac{A_i}{P_i^2}$, $4\pi\frac{\sum_i^n A_i}{(\sum_i^n P_i)^2}$, and $4\pi\frac{\sum_i^n A_i}{\sum_i^n P_i^2}$, where $i$ indexes the $n$ non-contiguous subregions of the district. Although the original Polsby--Popper score is bound to the range $[0,1]$, this is not true of the first of these alternatives. For a district with $n$ non-contiguous regions, the second alternative has a range of $[\frac{1}{n},1]$; this variant of the score penalizes districts for being non-contiguous. The final variant, which we use to calculate scores in this paper, yields a value of one if each non-contiguous region is a circle thereby acknowledging that non-contiguity may arise while encouraging each region of a district to be compact.

\begin{figure}[h!]
\centering
\includegraphics[width=0.3\columnwidth]{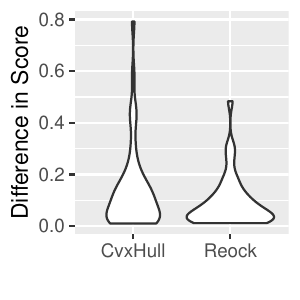}
\caption{Absolute value of differences in definitions of scores for districts of the 114th Congress. Shown are CvxHullPT vs.\ CvxHullPS and ReockPT vs.\ ReockPS. Districts are only shown if their score changed between definitions and they were part of a multi-district state, giving 47 data points for CvxHull and 38 for Reock. The data resolution was 1:500,000. \label{fig:definitiondiff}}
\end{figure}

Special attention should be given to non-contiguous districts to determine whether they result from natural features, legal requirements, or electoral engineering. In \autoref{fig:definitiondiff}, we calculate both the Convex Hull and Reock compactness scores for instances in which the polygons comprising a district are scored together versus separately, per \autoref{fig:interps}. Although the scores are nominally the same, a wide gap in values results from using the differing interpretations. This gap supports the need for precision in both language and implementation.

\subsection{Holes}

\begin{figure}[h!]
\centering
\includegraphics[width=2.8in]{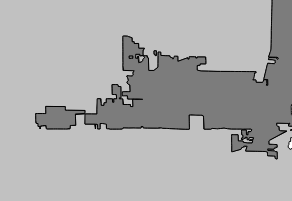}
\caption{Holes, islands, and narrow regions. The region shown is drawn from Wisconsin's Assembly Districts~\citep{wisconsin} and shows how poor digitization or subsequent simplification can lead to subtle data issues that are not visually apparent without significant magnification. Many borders are axis-aligned. Such alignment may reflect reality, but may also be an artifact arising from discretization of input data, demarcation choices, simplification algorithms, or even the visualization software. Regardless, axis alignment causes numerical issues in many simple geometric algorithms. 
\label{fig:border-artifacts}}
\end{figure}

Holes are relatively rare in districts, but many of the same considerations apply. The city of Racine, WI is noncontiguous due to annexations, as mentioned earlier. Placing the city within a single voting district required Wisconsin's legislature to draw the 61st State Assembly District in a way that creates both noncontiguity and holes (\autoref{fig:wisconsin}). Texas 18 very nearly surrounds the urban core of Houston and could, in a low-resolution dataset, contain a hole. Holes also appear as artifacts of the digitization process (\autoref{fig:border-artifacts}). For the 114th Congress 1:500,000 resolution data, four of 441 districts have holes as artifacts.

\subsection{Boundaries}

Districts are constrained by borders imposed by higher geopolitical units as well as by nature. Compactness scores that do not account for such constraints may assign low scores to a district that are not meaningful. The panhandles of Florida and Oklahoma, as well as Kentucky's border with the Ohio River (see \autoref{fig:polysimp}), contain electoral districts whose shape, at least in part, cannot be dictated by politics. The same is true of almost any coastal district since islands and peninsulas with their long perimeters must be included. Louisiana (\autoref{fig:polysimp}) exemplifies this challenge.

\begin{figure}[h!]
\centering
\includegraphics[width=2in]{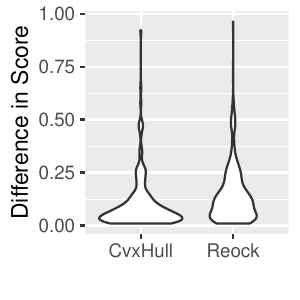}
\caption{Effects of constraining compactness measures using the boundaries of political superunits for the 144th Congress. The convex hull and, for the Reock score, the minimum bounding circle were cropped to state borders before being used to calculate scores. Only districts which were part of multi-district states and whose scores changed are shown: 215 for the convex hull and 320 for the Reock, of 441 total. District and state data were at 1:500,000 resolution. \label{fig:borders}}
\end{figure}

Some scores can be modified to account for this issue \citep{Azavea2006,Ansolabehere2016}. These can be marked with the suffix \textit{B} (borders accounted for). For example, in the case of the convex hull and Reock scores, if the hull or minimum bounding circle is intersected with a state polygon, the result is a better representation of what was possible and, therefore, a better indicator of whether gerrymandering took place. Taking boundaries into account in this way can have a considerable effect on compactness scores (\autoref{fig:borders}).

\begin{figure}[h!]
\centering
\includegraphics[width=2.5in]{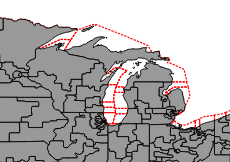}


\caption{Electoral districts of the 114th Congress including maritime regions. Two datasets of electoral districts are overlaid. The gray area depicts electoral district boundaries cropped to coastlines whereas the dashed red line indicates the full extent of the electoral districts. Note the growth of the district's areas and the relative smoothness of the perimeters.
Data was drawn from the US Census Bureau~\citep{shapefiles}; cropped data is from the Cartographic Boundaries dataset, e.g.\ \href{https://www.census.gov/geo/maps-data/data/cbf/cbf_cds.html}{\textit{cb\_2015\_us\_cd114\_rr.zip}}, whereas uncropped data is from the TIGER/Line dataset, e.g., \href{https://www.census.gov/geo/maps-data/data/tiger-line.html}{\textit{tl\_2015\_us\_cd114.shp}}. \label{fig:waterboundaries}}
\end{figure}

The boundaries of electoral districts, states, and countries may include large maritime regions, as shown in \autoref{fig:waterboundaries}. These regions are difficult or impossible to populate, except near shores, so their inclusion in compactness calculations may hide the effects of gerrymandering. Input data should be cropped to major coastlines to account for this, though doing so is not a panacea. Coastlines tend to be fractal and need to be measured in a way which is insensitive to this effect, as shown in \autoref{fig:koch}.

\begin{figure}[h!]
\centering
\includegraphics[width=2in]{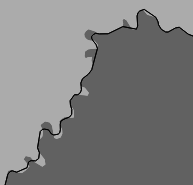}
\caption{Misaligned boundaries. Different sources of data place state and district boundaries in different places. The border shown here lies between Maryland 06 and West Virginia 01. The ``true boundary'' is drawn from data at 1:500,000 resolution and is shown by the transition between solid colors, while the black line shows the same boundary using 1:5,000,000 data. Differing data resolutions is only one instance in which a mismatch might occur: Shifts in data (as from projections), differing collection procedures, or deliberate manipulation are all possible as well. \label{fig:border-misalign}}
\end{figure}

\begin{figure}[h!]
\centering
\includegraphics[width=2in]{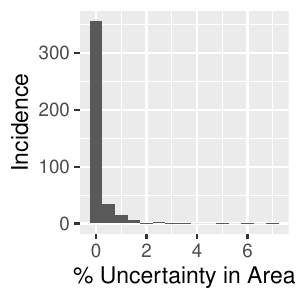}
\caption{Approximate percent uncertainty in area introduced by border misalignment.
Areas with especially high uncertainty are usually coastal where the lower resolution data introduce significant areas of water into a district. Using the data from \autoref{fig:border-misalign}, an exclusive-or on each district and state yielded areas of misalignment. Districts and states were shrunk and expanded to form border outlines which were intersected with the exclusive-or thereby limiting misalignment to border areas. The remaining area divided by the original, high-resolution areas gives the percentage. A few especially small districts ($<10$\,km$^2$) were culled from the analysis as this method made the entirety of the districts uncertain. \label{fig:misalign-effect}}
\end{figure}

As \autoref{fig:border-misalign} shows, boundary data, especially when drawn from disparate sources, may not always co-align. We attempted to quantify this effect by overlaying high-resolution district data with medium-resolution state data and found that the impact was usually small (see \autoref{fig:misalign-effect} for details). Problems can be avoided entirely by using data that is co-aligned, such as the data available from the U.S.\ Census.

\subsection{Projections}

\begin{figure}
\centering
\includegraphics[width=2.8in]{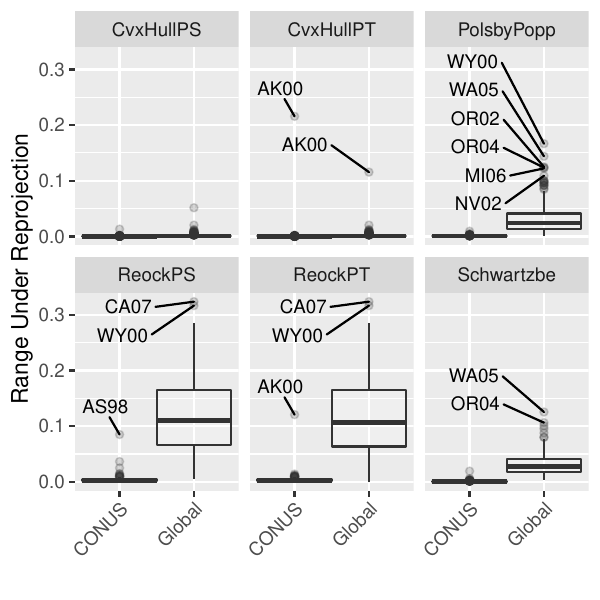}
\caption{Change in score between a locally-optimized projection and nationally- and globally-optimized projections for all electoral districts of the 114th Congress. Each district was projected into locally-fitted Lambert Conformal Conic and Albers Equal Area Conic projections; into conterminous US (CONUS)-fitted Albers Equal Area (EPSG:102003), Lambert Conformal Conic (EPSG:102004), and Equidistant Conic (EPSG:102005) projections; and into globally-fitted Mercator, Robinson, Molleweide, and Gall stereographic projections. For each district, the maximum range between any value in the local group and any value in the conterminous and global groups was calculated. For the conterminous projections, boxplot bodies appear as thin black lines indicating the bulk of districts experienced negligible change under different projections; in fact, the 99th quantile score across all districts was 0.009. The outlier is Alaska, for which a conterminous projection should never be used due to excessive distortion. If the entire United States, including Hawaii and Alaska, needs to be processed at once, Snyder's GS50 projection~\citep{snyder1984} is a good choice as it provides $<2\%$ scale distortion throughout this region. Data was at 1:500,000 resolution.\label{fig:projections}}
\end{figure}

Although scores are often defined as though districts exist on a plane, in reality they are wrapped around the curvature of the Earth and local topographical features. Several interpretations of scores are possible: Districts could be mapped to the plane using a projection designed to minimize distortion across an entire country, a subdivision of a country such as a state, or even the district itself. Alternatively, scores could be calculated on the sphere, WGS84 ellipsoid, or a similar body; we do not investigate this possibility here since it is used rarely in practice. As \autoref{fig:projections} shows, despite all the possibilities, compactness measures appear to be stable to \textit{reasonable} choices among \textit{localized} (country-scale) map projections used in practice. Alaska demonstrates what happens when an unreasonable choice is made: its score in a projection suitable for the conterminous United States differs from that of an Alaska-specific projection by up to 20\%.

Global projections, such as the standard Mercator, produce scores that differ markedly from local projections; therefore, global projections should not be used for calculating compactness scores---this includes the Web Mercator (EPSG:3857) projection, despite its ubiquitous use on the internet. Across all districts, scores, and projections, the absolute score difference between a district as measured in a locally-optimal projection versus a conterminous projection was less than 0.009 in 99\% of cases. The other 1\% of cases comprise districts such as Alaska and American Somoa, which are outside the region of interest for the conterminous projections. Given this observation, nation-sized projections---excluding outlying states and territories---are likely reasonable choices. Quantitatively, the conterminous Albers Equal Area (EPSG:102003) projection has a maximum scale distortion of 1.25\%~\citep{deetz1945}; this value can reasonably be taken as an upper limit on the acceptable distortion for any projection used to measure compactness.

\subsection{Topography}

\begin{figure}[h!]
\centering
\includegraphics[width=2in]{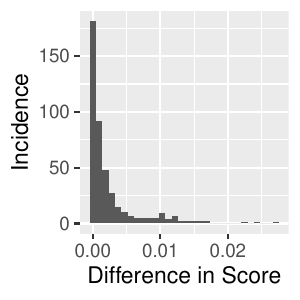}
\caption{Difference in Polsby--Popper scores when calculated on the plane versus with topography. Topography-inclusive area of districts was calculated using the 30\,m USGS National Elevation Dataset~\citep{ned}. Districts were cropped using 1:500,000 resolution boundaries from the US Census Bureau for the 114th Congress~\citep{shapefiles}. Surface area was calculated using RichDEM's implementation~\citep{richdem} of an algorithm by~\citet{jenness2004}. Perimeter was taken as the summed length of all the cells at the edge of a district and was constant with respect to topographic considerations. \label{fig:topography}}
\end{figure}

A different effect of mapping electoral districts to a plane is that topography, such as mountains, is left out of quantities such as area and perimeter. As a result, the true land area and overland distance between points is underestimated. Using the 30\,m USGS National Elevation Dataset~\citep{ned}, we calculated the surface area of districts using RichDEM's implementation~\citep{richdem} of an algorithm by \citet{jenness2004} and modeled perimeter as the summed length of all the raster elevation cells at the edge of a district.
The difference in Polsby--Popper scores between the topographic and non-topographic data was less than 0.03 for all districts, with 75\% of districts having deviations less than 0.005 (\autoref{fig:topography}). 

\subsection{Resolution}

\begin{figure}[h!]
\centering
\includegraphics[width=\textwidth]{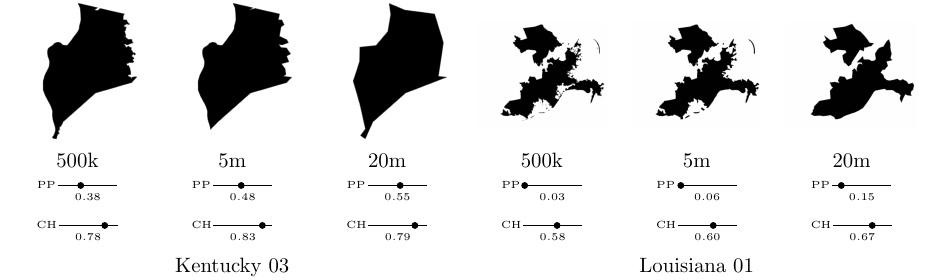}
\caption{Effect of polygon simplification on districts and their compactness scores. Districts from the 114th Congress are shown at 1:500,000 (500k), 1:5,000,000 (5m), and 1:20,000,000 (20m) resolution. Simplification was performed by the US Census Bureau using in-house algorithms that ensure border alignment. Here, PP stands for PolsbyPTAH while CH stands for CvxHullPT; note how these scores change with resolution. Kentucky 03 encompasses metropolitan Louisville and is bounded on the north by Kentucky's state border and the Ohio River. Louisiana 01 is bounded by the Mississippi Delta, divided by Louisiana 02, and includes unexpected parts of New Orleans. After simplification, the rough edges of Kentucky 03 disappear, as do entire bays and islands in Louisiana 01.
\label{fig:polysimp}}
\end{figure}

Resolution can be thought of as the density of points describing a boundary. \autoref{fig:polysimp} shows the same district at several resolutions. Lower resolutions obtained using standard simplification tools lead to simpler shapes often with shorter perimeters. The U.S.\ Census Bureau releases boundary data of Congressional Districts in four resolutions: full, 1:500k, 1:5M, and 1:20M~\citep{shapefiles}. The full-resolution data is available as ``TIGER/Line'' data whereas the other resolutions are available as ``Cartographic Boundary Shapefiles.'' At these resolutions, the perimeters of the districts of the 114th Congress are defined by an average of 8914, 1531, 322, and 70 points, respectively.


\begin{figure}[h!]
\centering
\includegraphics[width=3in]{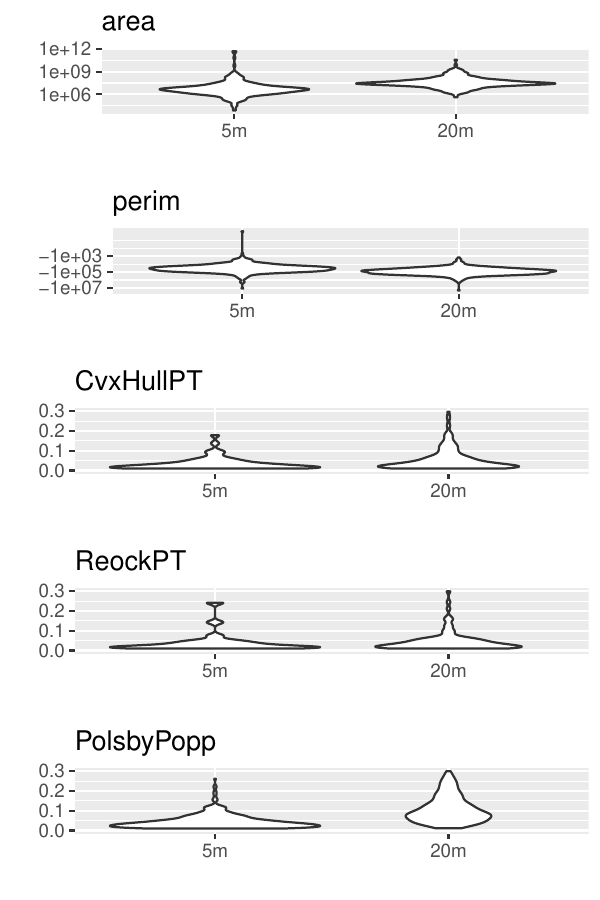}
\caption{Effect of resolution on compactness scores. Scores were calculated for districts from the 114th Congress at resolutions 1:500,000 (500k), 1:5,000,000 (5m), and 1:20,000,000 (20m). Score differences versus the 1:500,000 values are shown for those districts whose scores changed. Area and perimeter values are log-transformed. \label{fig:simp_together}} 
\end{figure}

\begin{figure}[h!]
\centering
\includegraphics[width=3.5in]{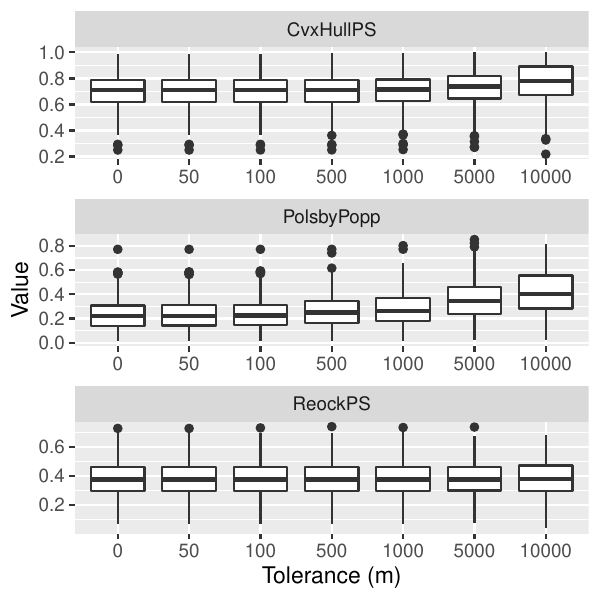}
\caption{Effect of polygon simplification on compactness scores. Districts from the 114th Congress were simplified by Shapely~\citep{shapely} using a topology-preserving algorithm from GEOS~\citep{geos} with the indicated tolerances. \label{fig:simp_indiv}}
\end{figure}

We find that the choice of resolution has a substantial impact on compactness scores (\autoref{fig:simp_together} and \ref{fig:simp_indiv}), with the popular Polsby--Popper score especially affected. This instability adds to a growing list of challenges for using the Polsby--Popper score in practice~\citep{Alexeev2017,chambers2010,deford2018total}. This suggests that lower-resolution data should be avoided, even if it could otherwise accelerate web and high-performance applications~\citep{tam2016}.

\begin{figure*}[h!]
\centering
\includegraphics[width=\textwidth]{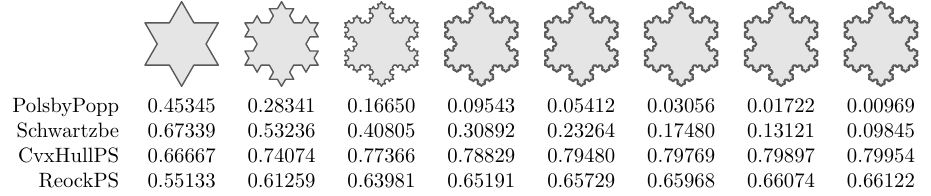}
\caption{The Koch Snowflake~\citep{koch1904} shown for its first 8 levels of resolution (the 0th level is omitted). At each resolution both the shape and boundary of the snowflake are visually similar, especially at higher resolutions; however, the levels have markedly different scores. For each increase in resolution the Polsby--Popper score decreases by 77\% and the Schwartzberg score by 33\%. After initial increases, the Convex Hull and Reock scores stabilize. \label{fig:koch}}
\end{figure*}

Since data may be supplied to users by outside sources, adversarial inputs are possible. Such inputs manipulate the data in ways which are sometimes hard to discern to alter measurement outcomes~\citep{Goodfellow2014}. A high-frequency wave applied to the boundary of a district may be visually imperceptible while introducing substantial alterations to a district's score. The Koch snowflake is an example of what an adversarial input might look: It has an arbitrarily-long perimeter surrounding a finite area (\autoref{fig:koch}). More practically, data may contain digitization or simplification artifacts that only become apparent under significant magnification, as shown in \autoref{fig:border-artifacts}.


\subsection{Choice}
\label{sec:choice}

If only one possible plan exists for a jurisdiction, that jurisdiction cannot be gerrymandered and should be excluded from analysis. In the Census Bureau data used here~\citep{shapefiles}, 13 states and territories, including Alaska, Delaware, and Vermont, had only one congressional district. No matter how oddly shaped these districts are, they are not gerrymandered.

\subsection{Floating-point Issues}
Computers generally store fractional values based on the IEEE754 specification using either the 32-bit single-precision type, which gives about 7 decimal places of precision, or the 64-bit double-precision type, which gives about 15 decimal places of precision. If geographic boundary data is in the form of decimal degrees of latitude and longitude, as is often the case, then storing such data in a 32-bit type is sufficient to resolve centimeter-scale features; storing such data in a 64-bit type provides nanometer-scale resolution. Thus, 32-bit single-precision types might be sufficient for storing geographic coordinates. However, performing mathematics on fractional numbers, especially 32-bit types, gives potentially erroneous results thanks to rounding and other effects~\citep{goldberg1991}.

We tested for floating-point instability by computing all of the scores mentioned here using both 32-bit and 64-bit IEE754 compliant types, with the latter taken as the ``true'' value. Compactness measured in these two systems differed by 
no more than 0.027\%.

\subsection{Ordering}

\begin{figure}[h!]
\centering
\includegraphics[width=2.5in]{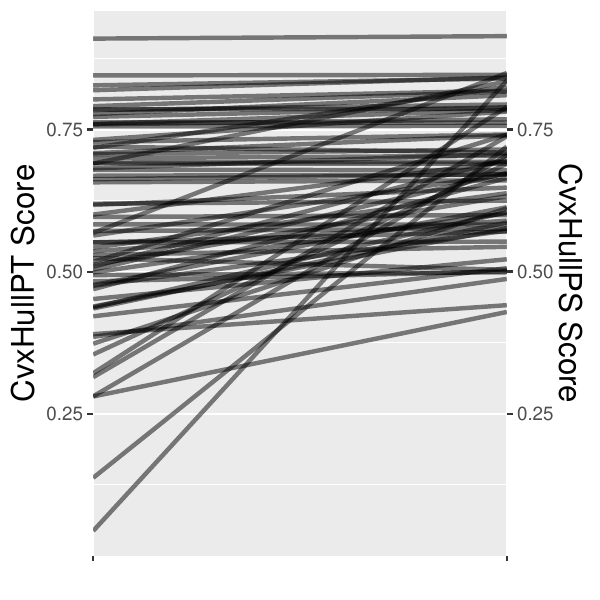}
\caption{Implementation affects ranking. Here, the compactness scores for the 114th Congress at 1:500,000 resolution are plotted for two different interpretations of the convex hull score. Only the 136 out of 441 districts whose score changed as a result of the differing interpretations are shown. \label{fig:order}}
\end{figure}

\begin{figure}[h!]
\centering
\includegraphics[width=3.5in]{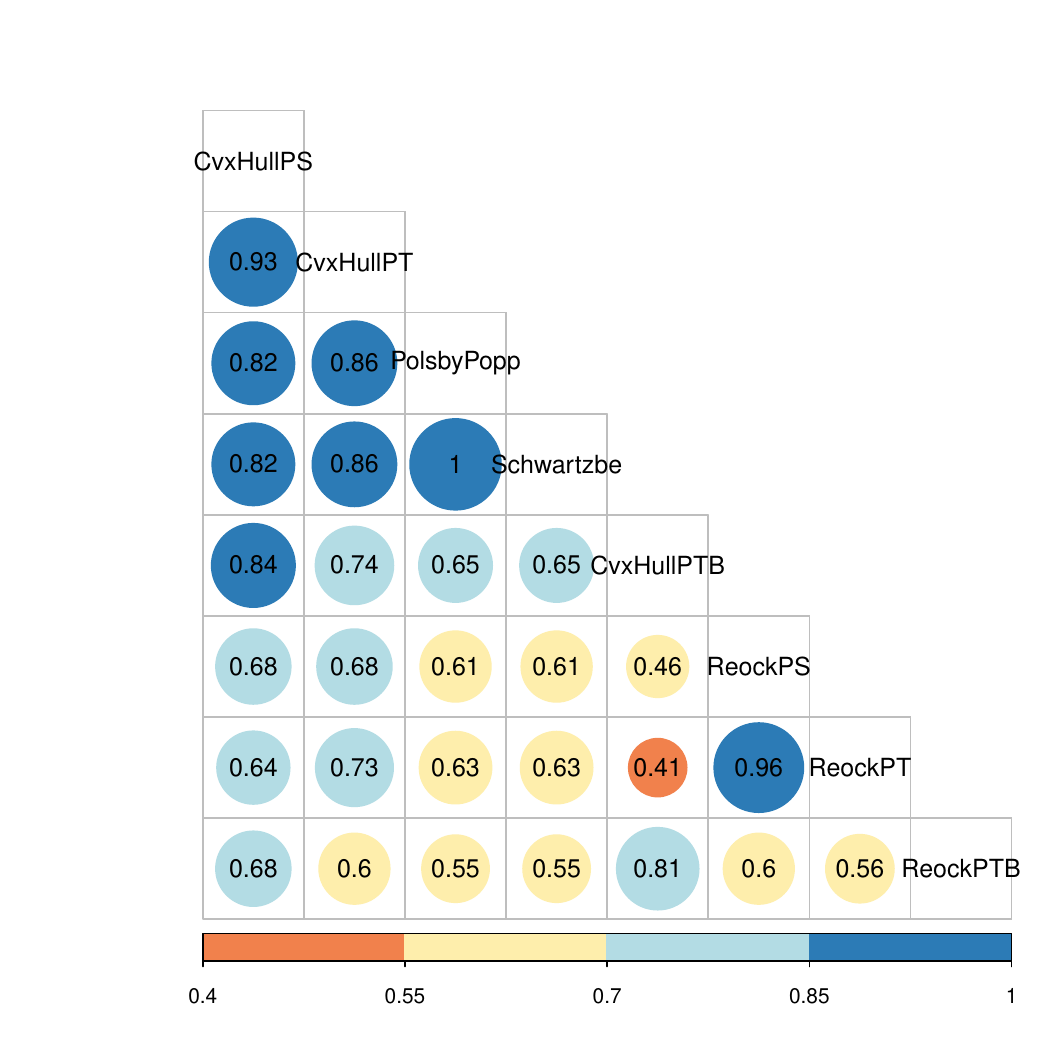}
\caption{Correlations of rankings. Rankings of compactness scores for the 114th Congress at 1:500,000 resolution are compared against each other using Spearman's rank correlation coefficient. A value of one indicates perfect agreement of relative rankings while a value of zero indicates no correlation. \label{fig:order_spear} }
\end{figure}

The foregoing considerations change not only the values of calculated scores, but also their relative ordering (\autoref{fig:order}). If ordering is quantified using Spearman's rank correlation coefficient (\autoref{fig:order_spear}), it is apparent that different scores give markedly different rankings. Thus, any ranking of districts by compactness is thoroughly tied to and arises from choices made in developing the scores. \autoref{fig:evil} explores this issue further, as described below.

This section listed many of the major decisions that must be made to measure compactness. These decisions may be made in good faith by people making measurements without awareness of their implications. They may also be made by adversarial actors seeking to affect the outcome of political decisions. The decisions are not independent of each other. In combination they provide more flexibility in outcomes than any one decision does by itself. We explore this below, in Section \ref{sec:results}.


\section{Results}
\label{sec:results}

A number of choices must be made to compute a compactness score. In addition to the choice of
(1)~compactness definition, we have shown that it is also important to consider how to handle
(2)~non-contiguous districts,
(3)~districts with holes,
(4)~political superunit boundaries,
(5)~map projections,
(6)~topography,
(7)~data resolution,
(8)~floating-point uncertainty,
and
(9)~whether alternative choices were possible in drawing a district's boundaries.

In combination, these choices provide unanticipated and undesirable flexibility. This flexibility can be abused. Different implementation choices applied to what is nominally the same score can lead to very different conclusions about the fairness of a districting plan.

To demonstrate this effect, we have selected ten U.S.\ Congressional Districts widely considered to be gerrymandered. For each district, we performed a grid search over a range of values for each implementation choice, thereby applying the full flexibility detailed in this paper. Similarly to electoral outlier analysis \citep{Ramachandran2018}, we were able to find sets of implementation decisions for which these districts' compactness scores are outliers when compared against the full distribution of district scores. We were also able to find sets of decisions which make these districts appear reasonable by locating them near the mean of the distribution. That is, we can exploit implementation flexibility to build seemingly reasonable arguments that these districts are not gerrymandered, as well as to build arguments that they are.

\begin{figure*}[h!]
\centering
\includegraphics[width=\textwidth]{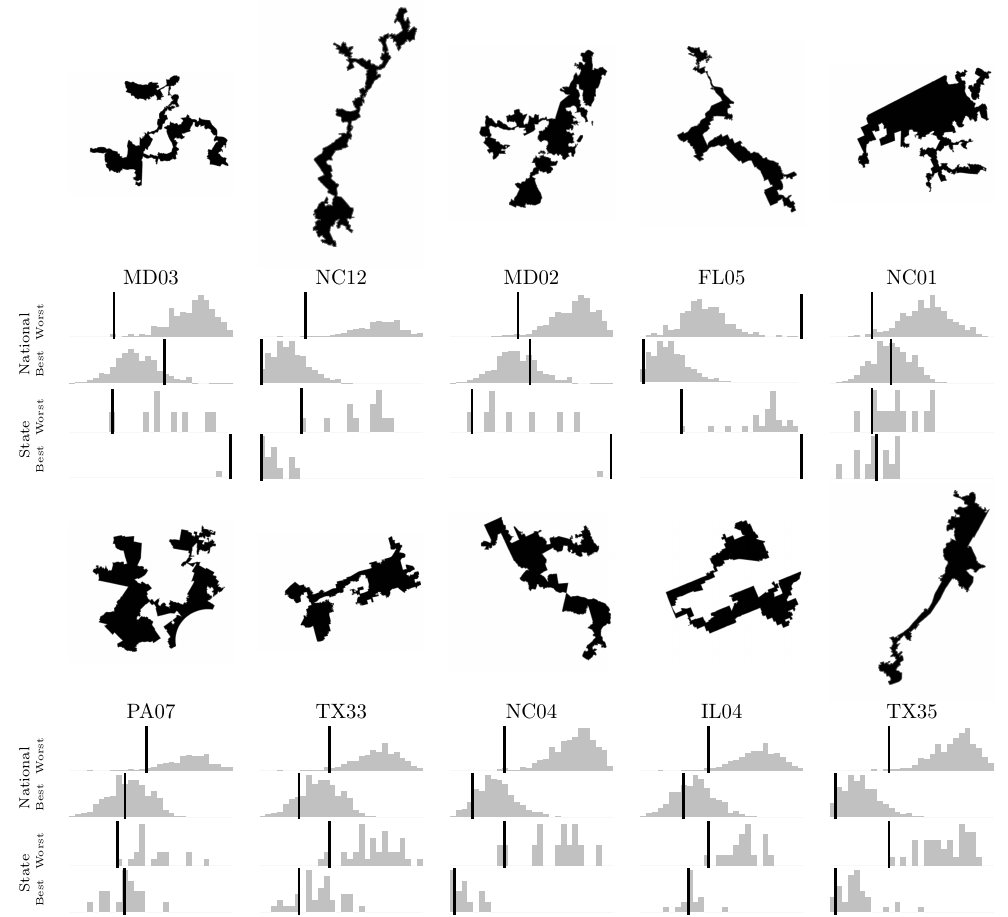}
\caption{Applied gerrymandering: abusing implementation flexibility. This figure shows several districts from the 114th Congress that appear incontrovertibly gerrymandered. We compare their compactness versus other districts in their state and nationally. The compactness scores of all the districts are shown in histograms with a black line indicating where the focal district falls in each distribution. Compactness ranges from 0 on the left-hand side of each histogram to 1 on the right. Scores for districts were generated by performing a grid search over a range of values for each implementation choice and choosing those values which minimized and maximized the difference between a districts' compactness score and the mean compactness score across all districts; that is, we found the choices which made each district look both the most gerrymandered as well as the most reasonable.
Further details for the figure are in \autoref{tbl:evil}. \label{fig:evil} }
\end{figure*}

\autoref{fig:evil} shows the effects of such adversarial choices of parameters. Considered against all districts nation-wide, in the case of NC01, IL04, and PA07, it was possible to move the districts from being obvious outliers to having middle-of-the-pack status. In other cases, such as NC12, NC04, and TX35, it was not possible to move the districts to the mean of the distribution, but they could still be moved considerably closer, potentially obfuscating their outlier status. Similar effects were true when districts were compared only against other districts in their states.

As \autoref{tbl:evil} shows, the optimizer does not need to use extreme settings to produce the desired results.
For example, TX33 appears most gerrymandered using the CvxHullPTB score at a 500\,m simplification tolerance in a locally-optimized Lambert conformal conic projection with all districts included in the distribution; it appears least gerrymandered using the ReockPT score with a 500\,m tolerance in a Gall projection with districts comprising an entire state excluded. A sensitivity analysis of the optimizer shows that the choice of score (e.g.\ Polsby--Popper, Reock) makes the greatest difference in the results, while the other choices all have similar effect sizes.

\subsection{Open Source Tools}

Of the many compactness scores discussed in the literature, some are better able to cope with the complexities discussed here than others. Many of the more robust metrics, however, are also difficult or impossible to calculate using commonly-available software. For instance, QGIS~\citep{qgis} includes the area of multipolygons as a built-in display field, convex hulls as a function three menu levels deep, and has no functionality to calculate the minimum bounding circles needed for Reock scores. 

To address this situation, we have released a family of open source packages which share a common library designed to efficiently, reproducibly, and correctly calculate a variety of compactness scores. The basis of this ecosystem is \texttt{compactnesslib},\footnote{\url{https://github.com/gerrymandr/compactnesslib}} a C++ library and associated command-line interface which ingests bulk or single data in a variety of formats and calculates compactness scores. The \texttt{python-mander} Python package\footnote{\url{https://github.com/gerrymandr/python-mander}} (available via pip\footnote{\url{https://pypi.python.org/pypi/mander}}) and the \texttt{mandeR} R package\footnote{\url{https://github.com/gerrymandr/mandeR}}
provide high-level interfaces to this library. In addition, a QGIS plugin\footnote{\url{https://github.com/gerrymandr/qgis-compactness}} provides GIS users an easy means of calculating scores~\citep{compactnesslib,python-mander,mandeR,qgismander}. This stack was utilized to produce the calculations in this paper: The complete source code for generating all the diagrams presented here is available at \texttt{\href{https://github.com/r-barnes/Barnes2018-compactness-implementation}{https://github.com/r-barnes/Barnes2018-compactness-implementation}}.

Though this software has the potential to improve the measurement of compactness as embodied by the scores we consider here, it cannot solve gerrymandering on its own: there are many ways to engineer districts each of which has its own flexibility. In this sense the software represents a model of the specificity, accessibility, and transparency necessary for any method of measuring gerrymandering or drawing districts.

\section{Discussion}
\label{sec:discussion}

\subsection{Best Practices}

Our results show the importance of clarity and transparency in the measurements used to evaluate potential voting districts. In general, a mathematical definition alone is not sufficient. Attention must be paid to data and algorithmic quality. As a model for the level of specificity needed to describe quantitative measures of districting plans, we suggest best practices for the calculation of compactness scores. These guidelines are the minimal set any expert would need to explicitly consider when evaluating compactness.
\begin{itemize}
\item \textbf{Scores.} Be explicit about what each variable in a compactness score means. Does area include holes? Is it constrained by political superunits? How should non-contiguous districts be handled? Score names should be distinct and informative. Appending a clarifying suffix to the name of a score (e.g.\ \textit{PTSHp}) informs readers about algorithmic details. See above for examples.
\item \textbf{Projections.} Scale distortion should be limited to only a few percent throughout the region of interest. Reasonable choices of national or local projections usually suffice.
\item \textbf{Resolution.} Use the best-available resolution from a trusted source. Simplified or down-scaled data give altered results. Alternatively, choose a score that is robust to changes in resolution, like hull-based scores or recent multiresolution measures \citep{deford2018total}. The U.S.\ Census Bureau produces reasonable data designed such that all borders that are at the same resolution align. Ideally, districting data should be drawn from a common, public, trusted, nonpartisan source.
\item \textbf{Border constraints.} Scores that do not explicitly account for constraints imposed by superunit boundaries leave out valuable information about what was possible in drawing a district. That is, they may unfairly penalize a district for having an odd shape when no other shape was possible. Use a score that accounts for superunit borders. Be sure that borders are cropped to features such as major coastlines.
\item \textbf{Choice.} Before doing statistics on a set of district plans, eliminate those districts that encompass an entire political superunit, as no other choices of shape were possible.
\item \textbf{Topography.} We have not found including topography in the calculation of area to be a significant source of variation, assuming the use of low-distortion map projections.
\item \textbf{Border coalignment.} Coalignment of borders is a concern, although the effect was small in our data. To avoid problems, datasets used in an analysis should always be at the same resolution and carefully coaligned during their creation. In the U.S., Census data satisfies these requirements.
\item \textbf{Floating-point considerations.} We have not found the choice of single- or double-precision floating-point representations to be a significant source of variation in our calculations.
\item \textbf{Transparency.} A compactness score should not be accepted and cannot be interpreted without knowing the steps that went into its creation. From a scientific standpoint, this consideration relates strongly to reproducibility: We cannot trust what we cannot reproduce. Therefore, documentation is needed down to the equation level, and the release of source code and data is critical~\citep{barnes2010,merali2010,ince2012}. FAIR principles should be adhered to~\citep{Wilkinson2016}.
\end{itemize}
More broadly, while compactness measures are attractive as quantitative means for analyzing districts, they are just a few of the many tools used to combat gerrymandering. Many other quantitative techniques and statistical measures are appearing in the academic literature and in practice. These can measure not only geometry, but also the effects of demography, voting patterns, and other relevant information.  Used together, these scores provide a more complete picture of the consequences of choosing one plan over another. However, they are subject to the same instabilities and potential for abuse identified above.  That is, the need for clearly-defined and well-understood quantitative criteria for assessing districts and plans extends far beyond geographical issues and should be a central point of discussion while considering new standards or legislation.  

\subsection{Policy Implications}

While the U.S.\ court system has declared that egregious gerrymandering is unconstitutional~\citep{scotus1986,usfed2016,scopenn2018}, they have not yet adopted a quantitative standard by which districts can be judged. In
Vieth v.\ Jubelirer (2004), the Supreme Court left open the possibility that a ``workable standard'' might exist~\citep{scotus2004}, but more recently the Court has shown skepticism saying that, ``partisan gerrymandering claims present political questions beyond the reach of the federal courts" (Rucho v.\ Common Cause, 2019). This paper demonstrates that any standard must be specified precisely and carefully, since differences in interpretation can have large effects on scores. Furthermore, our work demonstrates that even a well-specified standard may judge unreasonable districts as being reasonable (see \autoref{fig:evil}). Therefore, any legally-mandated standard of compactness should leave open the possibility of challenges. Moreover, given the implementation flexibility discussed here and its potential for abuse, courts should not accept quantitative arguments unless the code used to build those arguments is made publicly-accessible and inspected by experts.

\section{Coda}

Geometric compactness can be used as a tool to help detect and quantify gerrymandering. However, numerous engineering and implementation decisions must be made to calculate this quantity. The same is true of other such measurements. Whether used unintentionally or maliciously, this flexibility has strong bearing on the quality of measurements and can be leveraged to shape conclusions about the suitability of a districting plan. A measurement cannot be trusted unless complete information about its implementation is available.

Implementation flexibility, such as that discussed in this paper, has the potential to affect any method of measurement \citep{Gelman2013,Ioannidis2005}. Alternative ways of measuring the shapes of districts such as discrete geometries~\citep{Duchin2018} or multivalued scores~\citep{deford2018total} may be more resistant to such problems, but further investigations are needed to ensure that these methods are stable while still providing meaningful measurements. 

Beyond providing ``best practices'' for implementing compactness standards, we intend the open source software accompanying this paper as a first step toward fair and accurate compactness measurement, allowing scientists, politicians, and the public to evaluate plans using reproducible, mathematically well-founded, and computationally stable tools.

\section*{Funding}
US Department of Energy Computational Science Graduate Fellowship (DE-FG02-97ER25308) to RB; MIT Research Support Committee (``Structured Optimization for Geometric Problems'') to JS; Army Research Office (W911NF12-R-0011) to JS; National Science Foundation (IIS-1838071) to JS; Amazon Research Award to JS; National Science Foundation (ACI-1053575) to RB.

\section*{Acknowledgments}
The open source software described here had its genesis in the \textit{Geometry of Redistricting} workshop held at Tufts University August 7--11, 2017. John Connors helped develop the \texttt{mandeR} package. Max Gardner, Aaron Dennis, Daniel McGlone, and Ariel M'ndange-Pfupfu helped develop the \texttt{python-mander} package. Ariel M'ndange-Pfupfu and Vanessa Archambault helped develop the QGIS plugin.
Computation and data utilized XSEDE's Comet supercomputer~\citep{xsede}, which is supported by the NSF (Grant No.\ ACI-1053575).
Travel funding for RB and research support for JS was provided by a Prof.\ Amar G.\ Bose Research Grant and an Amazon Research Award. In-kind support was provided by Isaac B., Hannah J., Kelly K., Vivian L., and Jerry W.
\printbibliography


\begin{table*}[ht]
\centering
\includegraphics[width=0.7\textwidth]{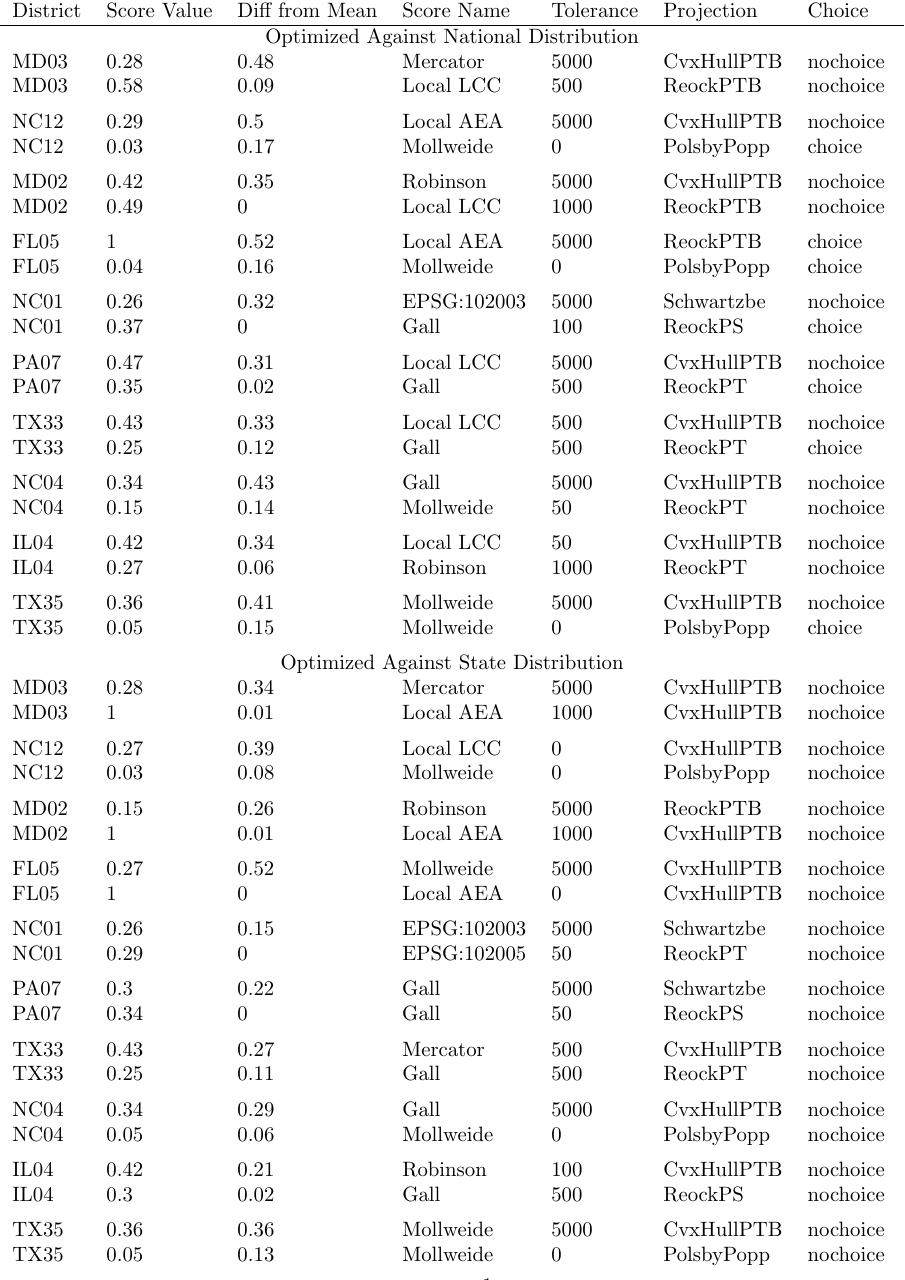}
\caption{Applied gerrymandering: abusing implementation flexibility. This table shows the choices made to produce the histograms shown in \autoref{fig:evil}. Recall that each of district which appeared incontrovertibly gerrymandered was paired with two histograms, one of which made the district's compactness score seem like an outlier and the other of which made it seem reasonable. The districts' scores are listed here, along with the absolute value of their difference from the mean of the distribution. The set of implementation choices made for each distribution is also shown: the compactness score, the simplification tolerance of the data, the map projection, and whether or not districts which comprised the entirety of their political superunit (districts in which a choice of boundaries was not possible) were included. \label{tbl:evil} }
\end{table*}

\end{document}